# Stabilizing stick-slip friction

Rosario Capozza[1], Shmuel M. Rubinstein[2], Itay Barel[1], Michael Urbakh[1], ,and Jay Fineberg[3]
[1]*School of Chemistry, Tel Aviv University, 69978 Tel Aviv, Israel*
[2]*Department of Physics, Harvard University, Cambridge, MA 02138, USA*
[3]*The Racah Institute of Physics, The Hebrew University of Jerusalem, Givat Ram, Jerusalem 91904, Israel*
(Dated: March 21, 2011)

Even the most regular stick-slip frictional sliding is always stochastic, with irregularity in both the intervals between slip events and the sizes of the associated stress drops. Applying small-amplitude oscillations to the shear force, we show, experimentally and theoretically, that the stick-slip periods synchronize. We further show that this phase-locking is related to the inhibition of slow rupture modes which forces a transition to fast rupture, providing a possible mechanism for observed remote triggering of earthquakes. Such manipulation of collective modes may be generally relevant to extended nonlinear systems driven near to criticality.

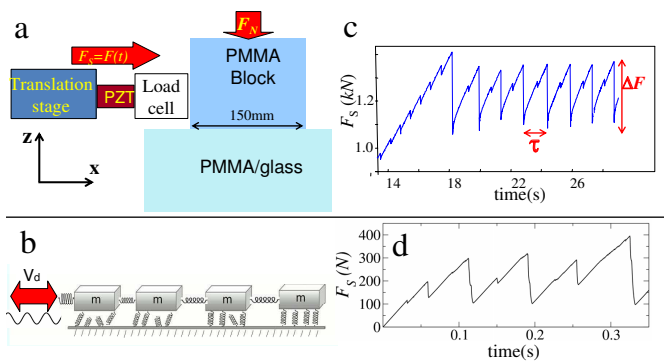

FIG. 1. Schematic views of the experimental system (a) and model (b). Typical stick-slip motion in experiments (c) and model (d) with no modulation of $F_s$. Parameter values used in model: $F_N = 4000N$, $K_d = 4 \times 10^7 N/m$, M=11.5kg, $V_0 = 10^{-4} m/s$, $K = 10^8 N/m$, $\tau_r = 0.005s$, $N_S = 20$, $N = 70$, $\gamma = 6462 s^{-1}$, $f_s = 1.37N$, $\Delta f_s = 6.9 \times 10^{-3} N$, $k = <k_i> = 5 \times 10^6 N/m$, $\Delta = 1\mu m$.

While the frictional motion of a single block is often considered to be described by wholly deterministic laws, a close examination reveals surprising variability in frictional processes. Stick-slip friction is one example. Although models [1] predict well-defined stick-slip frequencies, experiments reveal that intervals between successive stick-slip events have relatively broad distributions [2].

Let us consider two blocks, as in Fig.1a, that are pressed together with a normal force, $F_N$. When a shear force, $F_S$, is applied to the edge of one of the blocks, the onset of motion in this "simple" frictional system is surprisingly complex. The nonuniform stress profile produced by $F_S$ excites a sequence of rupture fronts created by successive failure of the ensemble of discrete contacts that forms the interface between the blocks. Initiating well before the onset of macroscopic motion, each rupture propagates from the loaded edge and arrests prior to traversing the entire interface [3–6]. Such avalanche-like collective motion occurs in many forced physical systems where numerous discrete degrees of freedom are spatially coupled [7], in the vicinity of a phase transition. In a frictional system, when each contact is near its rupture threshold, the onset of motion is mediated by three distinct types of collective modes; rapid subsonic and supersonic ruptures as well as "slow" rupture fronts [8, 9], nearly 2 orders of magnitude slower. Once initiated, the rupture velocities are coupled to the local ratio of shear/normal stress at the interface [9].

The nucleation of these ruptures is still far from understood. Recent experiments [9] have revealed that interfaces are locally much stronger than previously thought; sustaining local stress ratios a few times larger than the static friction coefficient without succumbing to motion. On the other hand, earthquakes were found to be triggered by small perturbations generated by either tidal forcing or other very remote earthquakes [10]. These questions motivated studies of slip onset in rock samples separated by a granular layer, upon application of sinusoidal perturbations to $F_S$ [2, 11, 12]. Extending previous work on the effects of oscillatory modulation of $F_N$ on reducing dynamic friction [13], the results suggested non-trivial dependencies on the phase [11] amplitude and frequency of the perturbation [2, 12].

Here we demonstrate, both in experiments and in a simple model, that the random intervals between stick-slip events can be stabilized by adding a low amplitude oscillatory component to $F_S$. Over a wide range of driving frequencies, a well-defined phase relation exists between a dimensionless variable characterizing the forcing function and the frictional onset. Moreover, we find that this phase-locking is related to a forced transition between slow to fast rupture modes.

Our experiments (Fig.1a) were performed on optically flat interfaces composed of two PMMA blocks, a slider and a base, that were roughened to about $1\mu m$ rms. The slider had $(x, y, z)$ dimensions of $(150, 6, 70) mm$ in the sliding, transverse and normal loading directions, respectively. The base blocks had $(x, y, z)$ dimensions of $(230, 30, 30) mm$. A constant and uniformly distributed normal force, $1000 < F_N < 5000N$ was imposed at the start of each experimental run. $F_S$ was applied to the slider's trailing edge via a stiff load cell (Kister 9602A) in series with a piezoelectric actuator (Piezomechanik Gmbh) and a translational stage moving at constant velocity, $V_0$. A ramped and modulated shear force, $K_d(V_0 t + \Delta \cos(2\pi t/T))$, was imposed, where $K_d \simeq 10^9 N/m$ is the system stiffness, $0.001 < T < 3s$ and $\Delta$ are the period and amplitude of sinusoidal oscillations generated by the actuator. The real contact area, $A(x,t)$, was continuously measured, as in [8].

We model the experiments (Fig.1b) by a slider of mass M interacting with an immobile rigid substrate. The slider is pushed from its trailing edge via a spring that is displaced at a velocity $V_d = V_0 - 2\pi/T\Delta \sin(2\pi t/T)$. The measured force exerted on the slider is $F_S = K_d(V_0 t + \Delta \cos(2\pi t/T)) - X_t)$, where $X_t$ is the position of the trailing edge. The slider is composed of $N$ rigid blocks coupled by springs of rigidity, $K_{int}$, so that $K_{int} = (N-1)K$, where $K$ is the slider rigidity. The friction between the slider and the immobile substrate, is described in terms of interactions between each block and the substrate through an array of surface contacts. Each contact is modeled as a spring of elastic constant $k_i$ connecting the block and the substrate, where $k_i = 1, 2, ... N_s$, and $N_s$ is the number of contacts between the block and the substrate. As long as a contact is intact, the contact's spring elongates or shortens with the velocity of the corresponding block, producing a force $F_i = k_i l_i(t)$ inhibiting the motion, where $l_i(t)$ is the spring length. A contact breaks when $f_i$ exceeds a threshold $f_{si}$. Contacts reattach in an unstressed state, after a delay time $t_r$ taken from the distribution $P(t_r) = e^{-t_r/\tau_r}/\tau_r$, where $\tau_r$ is a characteristic time of contact formation. The thresholds $f_{si}$ are chosen from a Gaussian distribution with a mean $f_s$ and a standard deviation $\Delta f_s$. We consider $f_{si}$ to be proportional to the area $A_i$ of the given contact, while the transverse rigidity $k_i$ is proportional to its size, $k_i \propto \sqrt{A_i}$. Therefore, the distributions of $k_i$ and $f_{si}$ are coupled by $k_i = k(f_{si}/f_s)^{1/2}$, where $k = <k_i>$. When a contact reattaches, new values of $f_{si}$ and $k_i$ are assigned to it. Artificial vibrations of the blocks are avoided by introducing a viscous damping force with a coefficient $\gamma$, $f_\gamma = -m\gamma \dot{x}_j$, where $x_j$ is the coordinate of the center of mass of the j-th block of mass $m = M/N$.

Both the experiments and model are in the stick-slip regime of friction (Fig.1). Previous studies [3, 6] showed that under these loading conditions large, system sized, stick-slip events are the culmination of a complex history of precursory rupture events that initiate at the system's trailing edge and arrest within the interface. The resulting slip generates discrete sequences of small sharp drops in the $F_S(t)$ curves (Fig.1) well below the peak values of $F_S(t)$.

When $F_S(t)$ is applied in the model, the shear stress accumulated at the trailing edge decays exponentially along the slider, with a corresponding deformation:

$$\Delta x_j = x_j - x_j^0 = \Delta x_1 \exp(-\sqrt{K_s/K_{int}}(j-1)) \quad (1)$$

where $x_j^0$ is the equilibrium position of block $j$, $\Delta x_1$ is the displacement of the first block at the trailing edge and $K_s = kN_S$. As $F_S$ increases, the stress grows until it reaches the threshold for rupture of surface contacts at the first block. The corresponding threshold displacement $\Delta x_{cr}$ of the blocks can be estimated as $\Delta x_{cr} \approx f_s/\langle k \rangle$. The experiments and simulations show that frictional sliding always has some stochasticity that appears as both uncertainties in the period between consecutive slip events (Fig.2) and irregularity in the size of the stress drops that follow each stick-slip event. The double peaked structure of the model's stick-slip interval distribution (Fig.2d) reflects contributions of stick-slip events with different numbers of precursors, mainly two and three in the simulations.

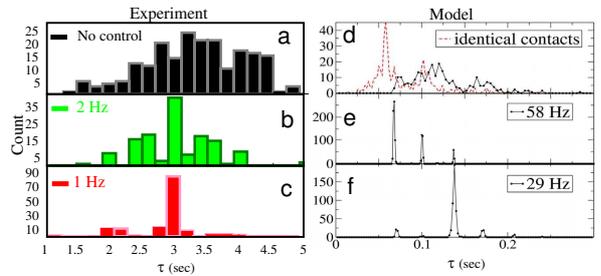

FIG. 2. Histograms of $\tau$ in experiments (left panels) and the model (right panels), when (a,d) no external control is imposed, and for small-amplitude sinusoidal perturbations with external frequencies of 2 Hz (b), 1Hz (c), 29Hz (e),58 Hz (f). The red dashed curve in (d) is a histogram calculated for identical surface contacts, when the system's nonlinearity is the only source of stochasticity. Upon imposition of the modulations, the intervals between stick-slip events become integer multiples of the perturbation period, $T$. The model parameter values used are as in Fig. 1.

What is the origin of the stochasticity of the frictional dynamics? The model has two sources of stochasticity: (i) a diversity of surface contacts characterized by distributions of rupture forces, stiffnesses and reattachment times, and (ii) nonlinearity of interactions between the driven spatially extended slider and the surface. Fig. 2d demonstrates that a broad distribution of stick times is retained even when all contacts are *identical*. Thus, the stochastic response is mainly due to the nonlinearity. The nonlinearity of the system leads to stochasticity only when the nonuniformly stressed region involves more than one block, i.e. $N >> \sqrt{K_{int}/K_s}$ (see Eq.(1)), which is also the condition for a proper description of a spatially extended elastic slider.

Once small harmonic perturbations are introduced to the shear loading, this picture changes significantly. Even relatively small perturbations can cause the interval, $\tau$, between successive stick-slip events to phase-lock to the perturbation frequency. This is clearly demonstrated when frequencies of 1 and 2Hz are imposed onto a uniform loading rate. As shown in the experimental histograms of $\tau$ in Fig.2, when a control signal is applied, the intervals between slip events are no longer randomly distributed around a mean value of $\tau_0 \simeq 3s$. Instead, the slipping interval, $\tau$, phase-locks to the driving frequency, $2\pi/T$ with $\tau$ attaining only integer values of $T$. The same synchronization also occurs in the simulations, as presented in Fig.2e,f. The model also provides partial synchronization of stress drops, which is not clearly evident in the experiments. Phase locking occurs, in both experiments and model, not only for $\tau$, but also for the intervals between suc-cessive precursors (see Supplementary materials).

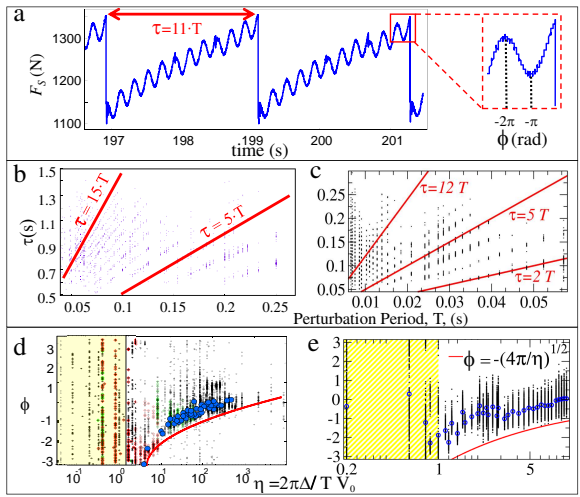

FIG. 3. Once phase-locking occurs, $\tau$ becomes an integer multiple, $n$, of the forcing period, $T$. (a) A typical experiment for large $n$ and $T \ll \tau_0$ ($\tau_0$ is the, unperturbed, mean stick-slip interval). Stick-slip intervals, $\tau$, as a function of $T$ for experiments (b) and the model (c) over a broad range of loading conditions (values of $\Delta$, $T$, and $V_0$). The data all fall on well-defined lines of $\tau = nT$ for different integer values of $n$. Examples are highlighted by solid lines. Plots of $\phi$ vs. $\eta$ for experiments (d) and model (e). Experiments: $1 < V_0 < 200 \mu m/sec$, $2 < \Delta < 100 \mu m$, $0.1 < 2\pi/T < 500 Hz$. Phase-locking is seen by the data clustering above a well-defined "back-bone". At low values of $\eta = \frac{2\pi\Delta}{TV_0}$ (shaded region), no phase-locking occurs and $-\pi < \phi < \pi$ occur with equal probability. As $\eta$ is increased, correlated events appear with the initial phase of $-\pi$ for small $\eta$ and saturate to zero for $\eta \gg 1$. Red curves in (d) and (e) correspond to $\phi_{min} \approx -2(\pi/\eta)^{1/2}$. Blue dots are averaged values of $\phi$ where phase locking occurred for given values of $\eta$.

In the results presented in Fig.2, $T$ was the same order as $\tau$. We find that phase-locking also occurs for control frequencies that are much higher than the typical stick-slip frequency, $\tau_0$. Figure 3 demonstrates that locking can occur at integer values $n$ such that $\tau = nT$. Essentially, $\tau$ adapts itself to the imposed value of $T$. Once $n$ is set, further increase (decrease) of $T$ during a given experiment will "drag" $\tau$ to respectively higher (lower) values, within the initial distribution of stick-slip intervals.

Let us now define the phase, $\phi$ of stick-slip motion relative to the forcing by the temporal shift, $\Delta t$, from the peak of the force modulation, $\Delta \cos(2\pi t/T)$, that succeeds the slip event (see inset in Fig.3). Thus, $\phi = 2\pi\Delta t/T$. When the synchronization demonstrated in Figs. 2 and 3 takes place, the system locks to a well defined-value of $\phi$ for each set of system parameters. A minimal possible value $\phi = \phi_{min}$ is set by the condition that the loading force associated with this phase should be higher than the preceding maximum of $F_S$ in the loading curve. This yields:

$$\phi_{min} + \eta \cos(\phi_{min}) \geq \Phi_{max} + \eta \cos(\Phi_{max}) \qquad (2)$$

where $\eta = \frac{2\pi\Delta}{TV_0}$ is a dimensionless parameter representing the ratio between the external harmonic control and the uniform background loading rate, and $\Phi_{max}$ is the phase corresponding to the maximum loading force that is defined by $\sin(\Phi_{max}) = 1/\eta$. For values $\eta < 1$, the loading force changes monotonically with time and harmonic oscillations do not influence the stick-slip pattern. For $\eta \gg 1$ Eq.2 predicts asymptotic behavior, $\phi_{min} \approx -2(\pi/\eta)^{1/2}$. The above predictions are verified by both the experiments and simulations presented in Fig.3d,e: (i) $\eta$ is indeed a relevant parameter that controls the frictional response to harmonic perturbations; (ii) A minimum value of $\eta \simeq 1$ exists, below which no phase-locking is observed; (iii) When control is applied a well-defined "backbone" exists, below which the onset of stick-slip motion will (nearly) never occur; (iv) This backbone is described by: $\phi \propto -\eta^{-1/2}$. The data for stick-slip events are strongly clustered above this curve.

It is important to note that $\eta$ may not be the only relevant control parameter in the system. The relation $\phi \propto -\eta^{-1/2}$ may be a necessary, but not sufficient condition for phase-locking. The values of $\phi$ for this analysis were determined by varying each parameter ($V_0$, $A$, and $T$) by 2 orders of magnitude or more (see caption of Fig.3). The phases of the vast majority of the non-phase-locked data are, however, above the backbone. This may imply that a weaker sort of stochastic phase-locking still persists. The $\phi \propto -\eta^{-1/2}$ scaling is, however, wholly consistent with previous measurements [11] in a granular system where strong phase locking was observed. Previous works [2, 11] also observed phase locking only for sufficiently large forcing amplitudes.

The behavior described in Fig.3 occurs for values of $\eta$ that span approximately 3 orders of magnitude, where phase-locking ranges from $1 < \eta < 1000$. In this sense, $\eta$ is certainly not a classic "small" parameter. One might ask whether at values of $\eta \simeq 1000$ we are still applying a small perturbation. The answer to this is decidedly yes. One way to see this is to compare the size of the applied perturbation over a single forcing period, $\Delta F_S = K_d \Delta$, to the value of $F_S^m$ needed to initiate stick-slip motion. Over the entire range of the data presented in Fig.3, we have $0.002 < \Delta F_S/F_S^m < 0.05$ in experiments and $0.05 < \Delta F_S/F_S^m < 0.1$ in simulations. Thus, despite its significant effect on the dynamics of stick-slip (strong locking is easily attained for a 1% forcing amplitude), the size of the perturbation is decidedly small. It should be noted that these perturbations have a negligible effect on the average friction force. Our simulations demonstrate that significant reduction of friction can be only achieved for very high oscillation amplitudes, $\Delta F_S/F_S^m > 0.2 - 0.3$.

To clarify the synchronization mechanism of stick-slip events, we compare in Fig.4b,d 2D maps of the fraction of attached contacts in the slider as functions of $x$ and $t$ in both the absence and presence of harmonic perturbations. The simulations show that the main effect of the perturbations on the detachment dynamics is the elimination of slow fronts. This effect is also observed experimentally in the contact area measurements (Fig.4a,c), by driving the system at sufficiently large frequencies to enable us to capture their effect on the front dynamics.

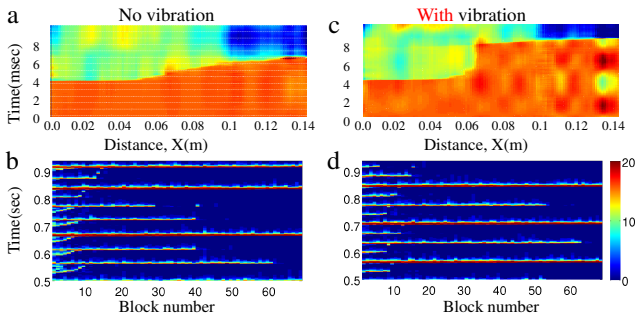

FIG. 4. Space-time maps of the measured real contact area (a,c) and fraction of attached contacts in the model (b,d) when no control is applied (left panels) and in the presence of oscillatory perturbations (right panels). Hotter (colder) colors indicate increased (decreased) contact area in (a,c). (b,d) The bars to the right of the maps set up a correspondence between the colors and the % of detached contacts.

What is taking place? Slow fronts arise [9] when the ratio of shear/normal stress is, locally, close to a critical, threshold value. An approximate 20% increase in this stress ratio is sufficient to trigger a transition to rapid rupture fronts. When applying stress to the system's trailing edge, its spatial distribution is highly nonuniform. Near the edge, as Eq.1 indicates, the effect of the modulating component of $F_S$ is therefore highly amplified and a 4-5% overall modulation could easily be amplified to a 20% modulation near the edge. Thus, near the critical stress ratio, slow fronts are arrested/inhibited for $\phi < \phi_{min}$ while rapid fronts may be triggered for $\phi > \phi_{min}$. If the stress modulations are sufficiently rapid, slow fronts couldn't propagate long before rapid rupture is nucleated. We surmise that this is what is occurring in Fig.4.

In summary, we have shown that the application of a small oscillatory perturbation has a significant effect on stick-slip dynamics, synchronizing the periods between consecutive slip events. We identified one of the relevant dimensionless parameters and showed how it functionally affects the locking phase. Based on both our experimental and theoretical results we proposed a model that explains the experimental observations and elucidates the mechanism for phase locking. This picture may bear relevance to observations of remote triggering of earthquakes. A fault which is susceptible to external triggering is most probably already close to criticality, perhaps approaching or within a state of slow, aseismic, rupture [14]. If the fault is subjected to non-uniform loading, a small amplitude, rapid stress oscillation radiated from a large faraway earthquake could be sufficient to trigger rapid, seismic, events, analogous to the one shown in Fig.4.

Our results may have interesting ramifications beyond frictional systems. Our system is but one example of a broad class of systems described by a large ensemble of nonlinearly coupled discrete variables that produce rich and complex collective behavior [7]. Thus, understanding how to excite and manipulate the collective modes involved in the stability of a simple rough frictional interface may provide insight into this general class of systems.

This work, as part of the ESF EUROCORES Program FANAS (ACOF), was supported by the Israel Science Foundation (1109/09). JF acknowledges support of the US-Israel Binational fund (grant no. 2006288), the James S. McDonnell Fund and the European Research Council (grant no. 267256). MU acknowledges support of the German-Israeli Project Cooperation Program (DIP).